# Transverse drag of slow light in moving atomic vapor

Y. Solomons*, C. Banerjee, S. Smartsev, J. Friedman, D. Eger, O. Firstenberg, N. Davidson

*Department of Physics of Complex Systems, Weizmann Institute of Science, Rehovot 76100, Israel*

*Corresponding author: yakov.solomons@weizmann.ac.il*

**ABSTRACT**

The Fresnel-Fizeau effect of transverse drag, in which the trajectory of a light beam changes due to transverse motion of the optical medium, is usually extremely small and hard to detect. We observe transverse drag in a moving hot-vapor cell, utilizing slow light due to electromagnetically induced transparency (EIT). The drag effect is enhanced by a factor $3.6 \cdot 10^5$, corresponding to the ratio between the light speed in vacuum and the group velocity under EIT conditions. We study the contribution of the thermal atomic motion, which is much faster than the mean medium velocity, and identify the regime where its effect on the transverse drag is negligible.

## Introduction

The idea of light drag starts in the 19th century, and it is closely related to the aether hypothesis. The first light drag experiment is due to Fizeau, who measured the propagation of light in water upstream and downstream [1]. The effect Fizeau measured was far weaker than expected by the aether theory, and it agrees with Fresnel's theory and with special relativity. In fact, this experiment greatly influenced Einstein when developing the theory of relativity [2]. Fizeau measured the longitudinal light drag, *i.e.*, due to longitudinal motion of the medium, which was confirmed by Michelson and others [3-5]. New applications, such as amplification of coherent light [6] and measurement of Fresnel drag for massive particles, such as neutrons [7], have been proposed.

In contrast, the transverse light drag is a change in the light path when it propagates perpendicular to the velocity of the optical medium. Traverse drag was first observed by Jones [8,9] by projecting light through a rotating glass disk. The measured drag was 6.17 nm for a rotation speed of 1500 rpm.

When the optical medium moves at a nonrelativistic transverse velocity $v$, the transverse drag leads to a displacement of the outgoing beam [10,11]

$$\Delta x = L \frac{v}{c} \left( \frac{c}{v_{\text{gr}}} - \frac{v_{\text{ph}}}{c} \right), \tag{1}$$

where $L$ is the medium length, and $v_{\text{gr}}$ and $v_{\text{ph}}$ are the group and phase velocities, respectively. The effect is intuitively understood as if the beam is 'dragged' by the moving material while maintaining its phase front and thus its original direction (Fig. 1a). Usually, $v_{\text{gr}} \approx v_{\text{ph}} \approx c$, so the drag effect is weak and extremely hard to detect.

Fortunately, light drag can be enhanced in media exhibiting strong dispersion and low group velocity [11-13]. This was verified for longitudinal drag with both hot and cold atoms [14-16]. For the transverse case, Franke-

Arnold *et al.* used self-pumped Ruby to enhance the drag in a rotating rod, using slow light from coherent population oscillations [17]. In fact, this served as proof that self-pumping contains a genuine slow-light component and not only pulse reshaping [18].

A mechanism prevalently used for realizing extremely slow light in atomic vapor is electromagnetically induced transparency (EIT) [19-21]. EIT occurs when a weak probe field and a strong control field couple two states of an atomic ground level to a common excited level. At the two-photon resonance, the probe experiences increased transparency, a near-unity refractive index (thus $v_{\text{ph}} \approx c$), and a steep dispersion, which leads to reduced group velocity $v_{\text{gr}} \ll c$. Under these conditions, Eq. (1) can be approximated as

$$\Delta x \approx \frac{Lv}{v_{\text{gr}}} = \tau v, \qquad (2)$$

where $\tau$ is the group delay of the probe in the medium.

Here we report on the transverse drag of slow light with an enhancement factor of $c/v_{\text{gr}} \approx 3.6 \cdot 10^5$, obtained by EIT in a hot atomic vapor. The medium moves with a slower linear velocity than in previous experiments. Taking care to measure both drag and delay we study the regime of validity of Eq. (2). Our experiments and models uncover corrections to Eq. (2) due to atomic thermal motion and diffusion.

## Experiment

The experimental schematics and atomic level structure are shown in Figs. 1b and 1c. We use a single diode laser at 795 nm matching the D1 transition in [87]Rb. The probe and control fields are formed by an electro-optic modulator tuned to resonance with the ground-level clock transition. The probe (control) couples the F=1 (F=2) hyperfine sublevel to the excited F'=1,2 sublevels. Both fields are detuned by $\Delta \approx 250$ MHz below F'=2. So that, the contributions from both excited sublevels render a symmetric EIT transmission line around the two-photon resonance $\delta = 0$; this choice emulates, as best as possible, a single Λ-system on resonance.

The probe and control, of orthogonal polarizations, exit a single-mode fiber, collimated to a waist radius of $w_c = 4$ mm, and sent to a spatial light modulator (SLM). The SLM is polarization-selective and affects only the probe. We use a grating phase-mask with a Gaussian aperture to reduce and vary the probe beam size. The widest probe used has a $w_p = 1.7$ mm waist radius, narrow enough to experience a nearly constant control intensity over its cross section. The total control power is 50 µW, and the maximal intensity of the probe is kept at 10%~20% of the control intensity.

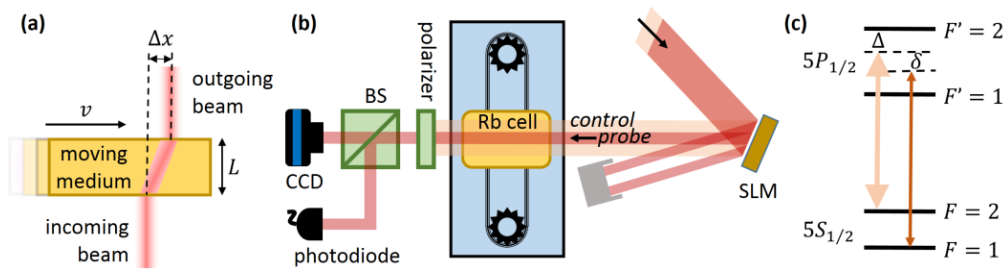

Fig 1. (a) Light is dragged transversely when traversing a moving medium. (b) A modulated narrowband laser forms the probe and control fields with orthogonal polarization. An SLM controls the probe size, and both beams are sent to a motorized cell which moves transversely back and forth. After the cell, a polarizer filters the control, and the probe is measured spatially and temporally. (c) The atomic transitions used in the experiment.

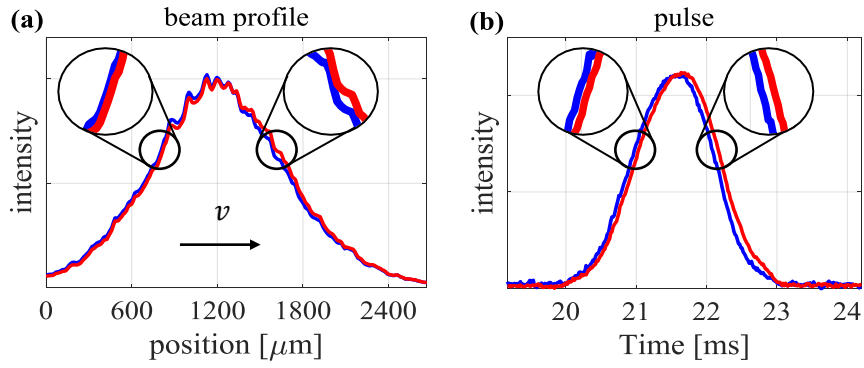

Fig 2. **Drag and delay of slow light in a medium moving transversely at $v = 200$ mm/s.** We compare the transmission on EIT resonance (red) to that off the EIT resonance (blue). (a) Beam profiles, obtained by vertical integration over the recorded images. The magnifiers display a small asymmetry between the right side of the beam (facing the direction of motion) and the left side, with the drag more prominent on the former. This is caused by the beam expansion due to atomic diffusion. (b) Temporal pulse shapes.

The Pyrex vapor cell (length $L = 7.5$ cm) contains a mixture of $^{87}$Rb vapor (temperature 55°C) and N$_2$ and Ar buffer-gases (10 and 90 Torr, respectively). At these conditions, the thermal motion of the rubidium atoms is diffusive, with a diffusion coefficient of $D \approx 110$ mm²/sec. We use a double-layer magnetic shield and solenoid coils to generate a small near-uniform magnetic field for lifting the degeneracy of the Zeeman sublevels. The cell assembly, weighing 890 grams, is mounted on a motorized stage (Thorlabs DDSM100), reaching maximal velocity $v = \pm 200$ mm/s. With optimized acceleration, the cell moves at a constant velocity when crossing the beam path.

After the cell, the control beam is filtered out by a polarizer, and the probe is split to two detectors: a CCD camera for measuring drag (imaging the exit facet of the cell) and a photodiode for measuring delay. We work on two-photon resonance ($\delta = 0$) and observe a delay of $\tau = 80 - 90$ μs, corresponding to group velocities of $v_g = 830 - 940$ m/s. The measured EIT linewidth is $2\gamma = 600$ Hz.

Figure 2 shows typical beam profiles and pulse shapes of the transmitted probe with the stage moving at $v = 200$ mm/s. We compare the transmission on EIT to that measured far from the EIT resonance. The mean drag is obtained from the shift of the center-of-mass of the two profiles (Fig. 2a). These profiles are always measured sequentially to reduce drift errors. Similarly, the delay is defined as the mean temporal shift between the pulses (Fig. 2b).

Results for different medium velocities are shown in Fig. 3. The drag follows the direction of motion and scales linearly with velocity. For a wide probe beam ($w_p = 1.4$ mm, Fig. 3a), for which diffusion effects are small, we observe a good agreement between the measured drag $\Delta x$ and the drag $v\tau$ expected from the measured delay $\tau$, over the whole tested velocity range. This agrees with the ideal picture of slow-light drag, represented by Eq. (2). For a narrower probe ($w_p = 0.42$ mm, Fig. 3b), for which diffusion effects are significant, the drag is still linear in velocity but with slope much smaller than $v\tau$, violating Eq. (2).

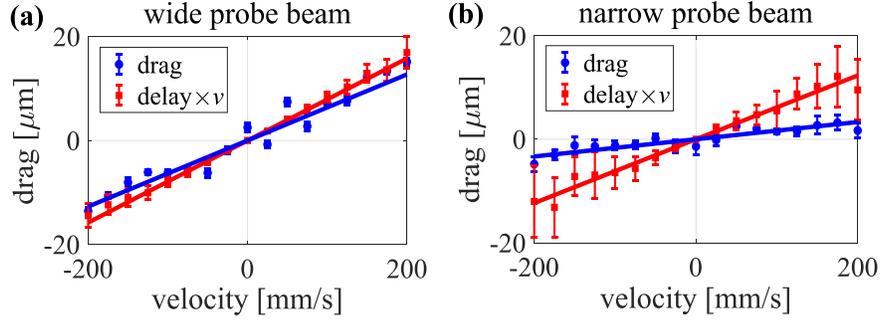

Fig. 3. **Transverse drag versus medium's velocity for two probe sizes**. The drag scales linearly with velocity in both cases. (a) For the wider probe ($w_p = 1.4$ mm), the measured drag (blue points) agrees with that expected from the measured delay (red points) within 20%. (b) For the narrower probe ($w_p = 0.42$ mm), the drag is substantially reduced, while the delay stays large. The lines are derived from a numerical model (blue line in Fig. 4), with the average delay being the only fit parameter. Data points are the means of 50 measurements, error bars are their standard deviations.

### Effect of diffusion for a finite probe

To understand the effect of thermal motion on the transverse drag, it is instructive to consider the optical linear susceptibility $\chi$ of the medium, as experienced by the probe. For simplicity, we assume degenerate probe and control fields, a plane-wave control, and a paraxial probe. Then $\chi = \chi(k_x, \delta)$ is a function of the transverse spatial frequency $k_x$ of the probe, and its detuning $\delta$ from EIT resonance. A probe field that is finite in space and time comprises a range of $k_x$ and $\delta$. Notably, the mathematics is the same for wave-packets in space and time, and a spatial displacement of a beam is akin to a temporal delay of a pulse,

$$\Delta x = \left(\frac{\pi L}{\lambda}\right)\frac{\partial \text{Re}[\chi]}{\partial k_x} \quad \text{and} \quad \tau = \left(\frac{\pi L}{\lambda}\right)\frac{\partial \text{Re}[\chi]}{\partial \delta}, \qquad (3)$$

where $\lambda$ is the optical wavelength.

If the medium moves transversely as a solid at velocity $v$, the dependence of $\chi$ on the wavevector mismatch between the probe and the control $k_x$ is due only to the residual (two-photon) Doppler shift $\delta \mapsto \delta + k_x v$. Substituting this relation in Eqs. (3) yields $\Delta x/\tau v = 1$, as in Eq. (2). However when the medium comprises gaseous atoms undergoing diffusive motion with a diffusion coefficient $D$, a wavevector mismatch $k_x \neq 0$ leads also to motional dephasing, broadening the EIT line by $Dk_x^2$ [22]. This broadening introduces an additional $k_x$-dependence of $\chi$, with no counterpart in terms of $\delta$, which violates Eq. (2). We quantify this violation by defining the drag coefficient $\beta = \Delta x/(\tau v)$, which is unity for solid media.

The effect of diffusion becomes prominent for narrow probe beams, when the motional broadening $Dk_x^2 \sim D/w_p^2$ is comparable to the bare EIT linewidth. Indeed for a narrow probe in Fig. 3b, we observe a 4-fold difference between the measured drag and that expected from the delay (drag coefficient $\beta = 0.25$). Figure 4 summarizes measurements for a range of probe beam sizes. The drag coefficient drops from nearly unity to nearly zero when the probe beam contracts.

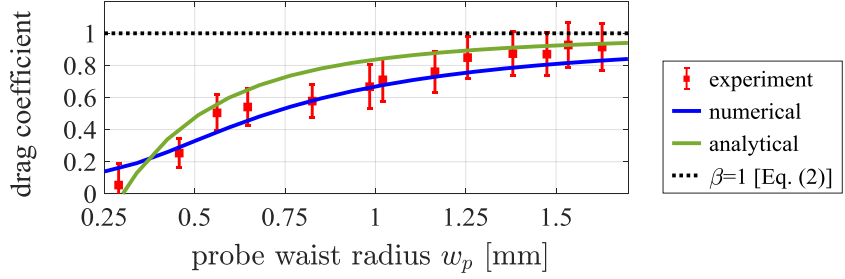

Fig. 4. **Deviation from Eq. (2) in terms of the drag coefficient $\beta = \Delta x/\tau v$.** Data obtained from measurements as in Fig. 3. Lines are models, using $D = 110$ mm$^2$/s, angular deviation $\theta = 3 \cdot 10^{-5}$ rad, optical depth OD $= 0.96$, and linewidth $2(\gamma_0 + \Omega^2/\Gamma) = 600$ Hz. The analytical model (green) is given by Eq. (7). The numerical model (blue) assumes slightly diverging probe with a radius of curvature of $R_c = 600$ mm.

## Probe and control of the same size

We now turn our attention to study the specific case of probe and control beams of the same shape and size, $w_\text{p} = w_\text{c}$. This arrangement is experimentally much simpler, as it does not require polarization-dependent shaping. Here the two fields are collimated together when exiting the fiber and sent directly to the cell. Besides simplicity, the main motivation is the reduced sensitivity to experimental imperfections, such as optical aberrations and angular deviation $\theta$ between the probe and control. When they travel a common optical path, they maintain perfect alignment ($\theta = 0$) and share identical aberrations, thus substantially reducing the sensitivity of two-photon processes, such as EIT, to imperfections.

Figure 5 presents measurements of transverse drag with two different beam sizes. In all experiments, the total control power is lowered by the ratio $[w_c/(4 \text{ mm})]^2$, so as to not change the maximal intensity. The drag is comparable to that achieved when $w_\text{p} < w_\text{c}$, despite the substantial variation of the EIT transmission across the probe beam due to the finite control beam (transmission drops from the center outwards). For $w_\text{p} = w_\text{c} = 1$ mm (Fig. 5a), we obtain a near-unity drag coefficient $\beta = 0.96$. This is higher than that obtained with a wide control beam ($\beta \approx 0.7$ for $w_\text{c} = 4$ mm, $w_\text{p} = 1$ mm, see Fig. 4), in which imperfections, such as small $\theta \neq 0$ and aberrations, reduce the attained drag. Here again, we observe a reduction in the drag coefficient for finite beams due to diffusion (Fig. 5b).

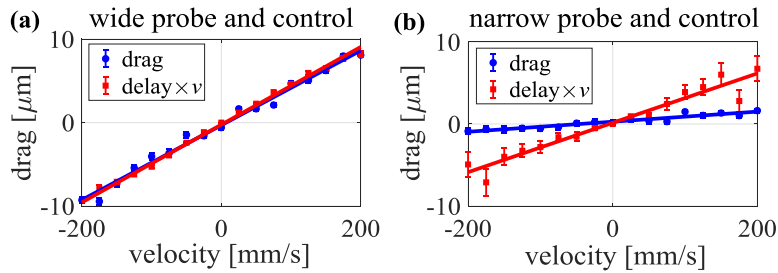

Fig. 5. **Transverse drag when the probe and control beams have the same size** (a) $w_\text{p} = w_\text{c} = 1$ mm, (b) $w_\text{p} = w_\text{c} = 0.17$ mm. The scattering of the data and their deviations from linearity are reduced compared to measurements with $w_\text{p} < w_\text{c}$ (compare to Fig. 3), due to the robustness of a common optical path as used here. This robustness also leads, for a probe wide enough (a), to an agreement between $\Delta x$ and $v\tau$ (i.e., $\beta = 1$).

## Theoretical model

To model transverse drag in a vapor medium, we derive the optical susceptibility following the formalism in Ref. [23]. We assume a Λ-system and a uniform control Rabi frequency $\Omega$. On top of the Maxwell-Boltzmann distribution of atomic velocities, we add to the model a mean drift velocity $v\hat{x}$. This introduces a residual Doppler shift $\delta_D(k_x) = -(k_x + 2\pi\theta/\lambda)v$ both for a finite probe beam (comprising nonzero transverse spatial frequencies $k_x \neq 0$) and for nonzero mean angular deviation from the control beam ($\theta \neq 0$). The linear susceptibility around $\Delta = 0$ is given by

$$\chi(k_x, \delta) = i\frac{\lambda}{\pi}\frac{\text{OD}}{2L}\left[1 - \frac{\Omega^2/\Gamma}{\gamma(k_x) - i\delta + i\delta_D(k_x)}\right], \tag{4}$$

where OD is the optical depth for the probe outside the EIT resonance, and $2\Gamma$ is the effective width of the one-photon resonance ($\Gamma = K^{-1}$ as defined in [23]) accounting for the homogeneous (pressure broadening) and inhomogeneous (Doppler broadening) contributions. The half width of the broaden EIT resonance is $\gamma(k_x) = \gamma_0 + \Omega^2/\Gamma + D(k_x + 2\pi\theta/\lambda)^2$, where $2\gamma_0$ is the bare linewidth, and the last term is the motional broadening in the Diffusion-Dicke regime [22].

The drag and delay can now be calculated using Eqs. (3). For the delay, we have

$$\frac{\pi L}{\lambda}\frac{\partial \chi}{\partial \delta} = \tau + i\eta\tau, \tag{5}$$

where the "imaginary delay" $\eta\tau$ causes a shift to the central frequency of the probe. We find

$$\tau = \frac{\text{OD}}{2}\frac{\Omega^2}{\Gamma}\frac{\gamma^2 - (\delta - \delta_D)^2}{[\gamma^2 + (\delta - \delta_D)^2]^2} \quad ; \quad \eta = \frac{2\gamma(\delta - \delta_D)}{\gamma^2 - (\delta - \delta_D)^2}. \tag{6}$$

Here $\gamma \equiv \gamma(k_x = 0)$, $\delta_D \equiv \delta_D(k_x = 0)$. For the drag, we find

$$\frac{\pi L}{\lambda}\frac{\partial \chi}{\partial k_x} = \Delta x + i\tau\left[v\eta + 2D\left(k_x + \frac{2\pi}{\lambda}\theta\right)\right].$$

The imaginary part leads to an angular deflection. The real part $\Delta x = \tau v - 2D(k_x + 2\pi\theta/\lambda)\eta\tau$ is the actual drag (deflection), which indeed differs from Eq. (2) when $D \neq 0$.

When considering a narrow probe beam, one needs to account for high-order corrections in the spatial frequencies $k_x$ comprising the field. As a rough approximation of the integral over $k_x$ for a probe with waist radius $w_p$, we sum the contribution of three terms: $k_x = 0$ (with weight 1/2) and $k_x = \pm w_p^{-1}$ (each with weight 1/4). Assuming $\theta \ll \lambda/w_p$, $\delta = 0$, and $|v| \ll w_p \times \gamma(w_p^{-1})$, we find

$$\beta \equiv \frac{\Delta x}{\tau v} \approx 1 - \frac{2}{1 + w_p^2(\gamma_0 + \Omega^2/\Gamma)/D}. \tag{7}$$

Indeed we find that the drag coefficient $\beta$ approaches 1 for wide probe beams and for slow diffusion. The green line in Fig. 4 is calculated using this model.

The blue lines in Figs. 3 and 4 are calculated numerically. We Fourier transform a Gaussian probe beam and calculate its propagation using the susceptibility in Eq. (4). The drag is determined from the center-of-mass shift after inverse Fourier transform. The delay is calculated by integrating over Eq. (6) in Fourier space.

The above model predicts for a uniform control that $\beta = 1$ whenever $D = 0$, i.e., that the drag $\Delta x$ is always equal $\tau v$ for a solid medium. It also predicts that both drag and delay saturate for high velocities, when the Doppler shift $k_x v$ becomes comparable to or larger than the EIT linewidth. To check these two predictions also for a finite control beam, we generalize the slow-light propagation equations of Ref. [23] by adding a transverse drift to the Boltzmann distribution of atomic velocities. Numerical solutions of these equations with a Gaussian control beam $\Omega(x) = \Omega e^{-x^2/w_c^2}$ and with $D = 0$, using a Laplace transform in space, are shown in Fig. 6. We find that $\beta = 1$ also for a finite control and that, for small velocities, the drag and delay do not differ from those with a uniform control. They saturate together for large velocities, with the saturation more pronounced for a finite control. The latter is a consequence of the reduced transmission outside the control beam, in effect preventing the probe beam from leaving it.

## Summary and outlook

We employed a moving cell of hot Rb vapor under EIT conditions to study the transverse drag of slowly-propagating Gaussian beams. The drag effect is clearly evident, despite the rapid thermal motion of the atoms. The transverse displacement due to drag depends linearly on the medium velocity, with a slope $\times\, 3.6 \cdot 10^5$ larger than that obtained for non-slow light. For probe beams wide enough, the simple relation $\Delta x = \tau v$ is preserved, even for finite control beams. For narrower probes, the drag is suppressed, $\Delta x < \tau v$, due to the thermal atomic motion. Our model, which takes into account the diffusion of the atoms, quantitatively explains all measurements with both approximate analytic expressions and exact numerical solutions.

In a related experiment, we have also observed the same transverse drag effect for a static atomic cell and transversely moving slow-light beam. The relation between these two complementary experimental schemes have been explored earlier [24, 25] and, for our medium velocities, the effects should indeed be identical.

Our work has both fundamental and practical significance. Understanding how a light beam is affected by complicated and arbitrary velocity fields has been a fundamental physical question for centuries, and the slow-light enhancement provided by our setup can enable such measurements under well-controlled laboratory conditions. This understanding and enhanced sensitivity can also be useful for sensing arbitrary velocity fields for practical applications, from inertial sensing and navigation, through precise measurements of gas flow and study of turbulence, to new types of optical image processing.

**Funding.**

This work was supported by the Pazy Foundation, and the Israel Science Foundation.

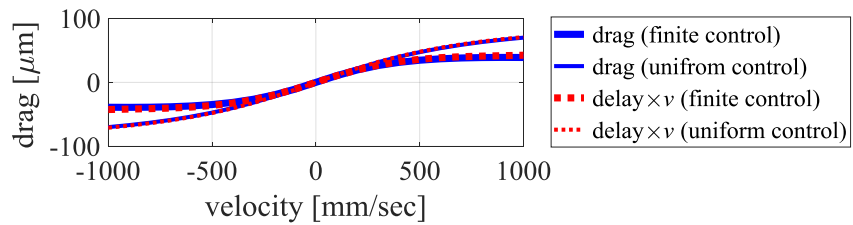

Fig. 6. **Numerical simulation of drag and delay for $w_c = w_p$ (finite control) and $w_c \gg w_p$ (uniform control), with $D = 0$.** Conditions correspond to the experiment with $w_p = 1$ mm. The saturation of the drag with increasing velocity is more pronounced for a finite control.